\documentclass[12pt]{article}
\newcommand{\Comment}[1]{{}}
\usepackage{amsfonts,amsthm,amsmath,amssymb,slashed}
\usepackage[textwidth = 430 pt, textheight = 630 pt]{geometry}

\Comment{\usepackage{color}
\definecolor{MyDarkBlue}{rgb}{0.15,0.15,0.45}
\usepackage[linktocpage=true]{hyperref}
\hypersetup{
colorlinks=true,
citecolor=MyDarkBlue,
linkcolor=MyDarkBlue,
urlcolor=MyDarkBlue,
pdfauthor={Horatiu Nastase and Jacob Sonnenschein},
pdftitle={The title},
pdfsubject={hep-th}
}

\usepackage[numbers,sort&compress]{natbib}
\usepackage{hypernat}}
\usepackage{graphicx}

\newcommand\ignore[1]{}
\def\one{{\,\hbox{1\kern-.8mm l}}}

\def\a{\alpha}\def\b{\beta}

\def\d{\partial}

\def \pa {\partial}

\newcommand{\Cset}{{\,\,{{{^{_{\pmb{\mid}}}}\kern-.45em{\mathrm C}}}}}

\newcommand{\be}{\begin{equation}}
\newcommand{\bea}{\begin{eqnarray}}

\newcommand{\ee}{\end{equation}}
\newcommand{\eea}{\end{eqnarray}}

\begin{document}

\renewcommand{\thefootnote}{\fnsymbol{footnote}}

\makeatletter
\@addtoreset{equation}{section}
\makeatother
\renewcommand{\theequation}{\thesection.\arabic{equation}}

\rightline{}
\rightline{}
   \vspace{1.8truecm}


\vspace{10pt}


\begin{center}
{\LARGE \bf{\sc Towards brane-antibrane inflation in type $II_A$: The holographic MQCD model}}
\end{center}
 \vspace{1truecm}
\thispagestyle{empty} \centerline{
{\large \bf {\sc Horatiu Nastase${}^{a,}$}}\footnote{E-mail address: \Comment{\href{mailto:nastase@ift.unesp.br}}{\tt
    nastase@ift.unesp.br}}
{\bf{\sc and}}
 {\large \bf {\sc Jacob Sonnenschein${}^{b,}$}}\footnote{E-mail address: \Comment{\href{mailto:cobi@post.tau.ac.il}}{\tt cobi@post.tau.ac.il}}
                                                           }

\vspace{1cm}

\vspace{.8cm}
\centerline{{\it ${}^a$
Instituto de F\'{i}sica Te\'{o}rica, UNESP-Universidade Estadual Paulista}} \centerline{{\it
R. Dr. Bento T. Ferraz 271, Bl. II, Sao Paulo 01140-070, SP, Brazil}}

\centerline{{\it ${}^b$
School of Physics and Astronomy,}}
 \centerline{{\it The Raymond and Beverly Sackler Faculty of Exact Sciences, }} \centerline{{\it Tel Aviv University, Ramat Aviv 69978, Israel}}

\vspace{1.0truecm}

\thispagestyle{empty}

\centerline{\sc Abstract}

\vspace{.4truecm}

\begin{center}
\begin{minipage}[c]{380pt}
{\noindent We describe type $II_A$ cosmological brane inflation scenarios based on the holographic MQCD model of Aharony et al \cite{Aharony:2010mi}.
The scenarios can be related via    T-duality  to the type $II_B$ KKLMMT  model \cite{Kachru:2003sx}.
They
describe a probe brane configuration of $p$ D4 branes stretching between an $NS5$  and $NS5'$ branes
in the holographic background of large $N$ D4 branes. The resulting cosmological models have a Wick-rotated D4-brane metric, with transverse
dimensions compactified, and a spiralling brane with flux $p$. In one model, the background has a small nonextremality, and the inflaton is
provided by the position of a ``sliding" D4-brane, and in  the other, the background is supersymmetric, but with a sliding anti-D4-brane.
We obtain  good and generic inflationary models, though several unknowns remain, in particular about subleading corrections.
The usual caveat of volume stabilization generically spoiling slow-roll still applies.
}
\end{minipage}
\end{center}

\vspace{.5cm}

\setcounter{page}{0}
\setcounter{tocdepth}{2}

\newpage

\renewcommand{\thefootnote}{\arabic{footnote}}
\setcounter{footnote}{0}

\linespread{1.1}
\parskip 4pt


\section{Introduction}

The inflationary scenario is by far the most successful in describing cosmological observations. However, embedding it in string theory
with sufficient generality has proven to be  quite difficult. One possibility to achieve it  is brane-antibrane inflation, but usually one has to arrange for
fine-tuned initial condition to obtain good slow-roll inflation \cite{Burgess:2001fx,Alexander:2001ks}. Another issue is obtaining a de Sitter
vacuum in string theory in a controllable way, since a supersymmetric vacuum will be Anti-de Sitter. In \cite{Kachru:2003aw}, such a
scenario was proposed, and since then other examples have emerged. This idea  was then used in \cite{Kachru:2003sx} to construct a model of brane-antibrane
inflation in type IIB that satisfies both slow-roll and generality (absence of fine-tuning),
though it was found that generically imposing stabilization of the
last modulus (volume) spoils slow-roll. In view of this, it is useful to construct other models of inflation, even if they suffer from one
of the above mentioned problems.

The Kachru-Kallosh-Linde-Trivedi (KKLT) model \cite{Kachru:2003aw} is a type IIB model obtained by compactifying the
Klebanov-Strassler (KS) solution \cite{Klebanov:2000hb}, i.e.
cutting off the KS cigar, a deformed cone with a  base $T^{1,1}$ (a 5 dimensional space, topologically $S^2\times S^3$)
and radial direction $r$, at a certain value of
$r$ and gluing a $CY_3$. Because of the ${\cal N}=1$ supersymmetry of the KS solution, the gluing procedure (compactification) occurs smoothly, without
extra energy needed at the junction. It is assumed that the dilaton and all the complex structure moduli are stabilized, and the only modulus
left is the volume modulus. The background solution is obtained from a large number $N$ of D3-branes, with fluxes and other branes (D7-branes,
Euclidean D3-branes) added in a supersymmetric configuration. An anti-D3-brane is added at the tip of the KS solution, breaking supersymmetry,
and lifting the AdS minimum to a dS minimum.

Then, Kachru-Kallosh-Linde-Maldacena-McAllister-Trivedi (KKLMMT) \cite{Kachru:2003sx}
considered a modification of the KKLT model, where there is an extra D3-brane sliding towards the anti-D3-brane
at $r_{min}$ in the compactified KS geometry. One obtains a model of D-brane anti-D-brane inflation, where the inflaton is the separation between
the brane and the antibrane, though it is found that generically requiring stabilization of the volume modulus  spoils the nice features of the
brane-anti-brane potential.

In the following we will search for a IIA counterpart to the IIB KKLMMT model, looking for inflation with the inflaton = position of a  probe brane.
The same caveat as for KKLMMT  applies to our model, namely requiring stabilization of the volume modulus will spoil the 
flatness of the potential. In this paper we will not discuss how to deal with this problem.

\begin{figure}[htp]
\begin{center}
\includegraphics[width= 80mm]{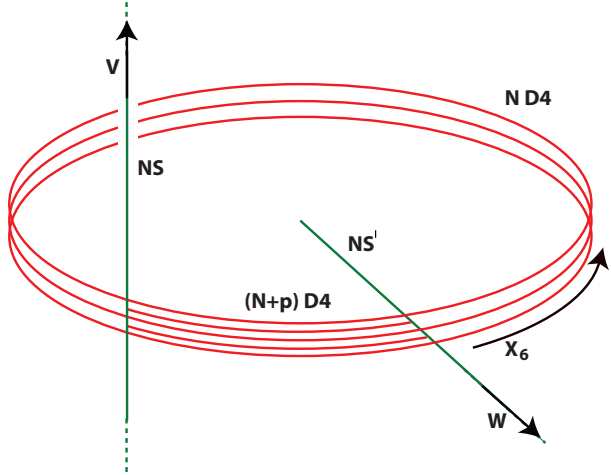}\end{center}
\caption{$N$ $D4$-branes that wrap the $x_6$ circle, and $p$
that are stretched between the $NS$ and $NS'$-branes.\label{compactified}}
\end{figure}

\begin{figure}[htp]
\begin{center}
\includegraphics[width= 60mm]{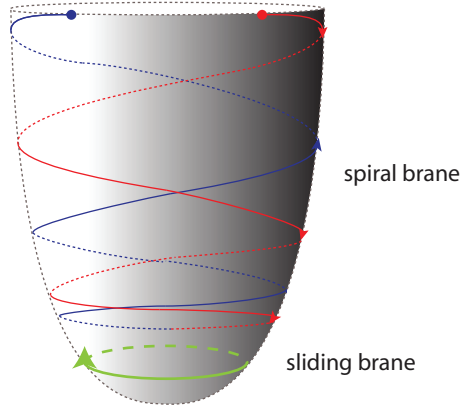}\end{center}
\caption{  The spiraling profile over the cigar background.
 We use a red line for the ``downward"  spiralling brane, a blue one for the climbing one and a green one for the sliding brane.
\label{cascading}}
\end{figure}

In this present paper we analyze inflation scenarios based on type $II_A$ models. The models are based on two versions of holographic
MQCD  that were proposed in  \cite{Aharony:2010mi}. These models consist of an uplift to M theory of a type $II_A$  brane configuration depicted
in figure  (\ref{compactified}). In both models  a world-volume  coordinate  of the D4 branes $x_6$  is compactified. In one case, the extremal
case, supersymmetry is unbroken and the geometry in the radial  and $x_6$ direction
is that of a cylinder (see figure (\ref{cigarspiral})). In the other case, the near-extremal one, supersymmetry is slightly broken by a small
nonextremality parameter, yielding a cigar-like geometry of figure (\ref {cascading}) (in the limit of zero non-extremality, we recover the
supersymmetric case).

In both cases the form of the ``probe p D4 brane plus the NS and NS' branes" take the form of a spiralling brane descending up to a certain radial value and then ascending back (see figures \ref {cascading}  and  \ref{cigarspiral}), but we will find that its effect on the inflaton potential is
a small correction, so we can just use the supersymmetric or near-supersymmetric backgrounds of  above without the spirals.
Our first inflation scenario is based on the latter case where a sliding D4 brane is added and its location along the radial direction plays the
role of the inflaton.  In the second scenario  the  inflaton is the location of anti-D4 brane in the supersymmetric background.

We should mention that, despite the fact that for the   supersymmetric  case the   background  we use  is  related  in the limit of a shrinking radius by T-duality to  the KS background,  the mechanism of generating  the inflaton potential cannot be obtained directly from the KKLMMT one, and  hence the scenario described here is a novel one.

The paper is organized as follows. In section 2 we will describe the set-up T-dual to KKLMMT, of a Wick-rotated near-extremal D4-brane with a
spiralling 5-brane added, and the position of a moving D4-brane providing the inflaton. Adding the spiral makes a small correction to the
potential. In section 3 we describe an alternative model, with an extremal D4-brane background and a moving anti-D4-brane. In section 4 we analyze
the resulting cosmology of these models, and in section 5 we conclude.

\section{A model based on a sliding D brane in a non-supersymmetric (cigar) background}

\subsection{The basic set-up}

If we T-dualize the KS solution to type IIA on a coordinate $x_6$, we obtain a solution made up of a large number $N$ of D4-branes wrapping $x_6$.
In \cite{Aharony:2010mi}, the solution T-dual to KS was described by a background of $N$ D4-branes, lifted to M theory as M5-branes, in which
one has other (probe) $p$ D4-branes stretched between an  NS5-brane and an NS5' brane.

 The $p$ probe D4-branes and 2 NS5-branes lift to M theory as a single 5-brane, winding
around $x_6$. In the cylinder formed by $x_6$ and the transverse partial radial coordinate $u$, the resulting 5-brane spirals down to an $u_{min}$, and
then goes back up. A particular limit of the background of $N$ D4-branes with the spiralling brane embedded was shown to be equivalent to the
KS construction for the T-dual type IIB.

The 5 coordinates transverse to the $N$ D4-branes are described by 2 complex variables $v$ and $w$, with
the (partial) radial coordinate $u$, $u^2=|v|^2+|w|^2$ and another one $x_7$, with the overall transverse coordinate $r$, $r^2=u^2+x_7^2$.
Therefore the transverse space is  sort of a semi-infinite cigar with radial coordinate $r$ and base $S^4$.

In the KKLT construction there is  also an anti-D3 brane at the tip of the KS solution, breaking susy in a controllable way. In the model we
study here, we choose to make the nonsusy perturbation a part of the background, by taking near-extremal D4-branes. More precisely, since we are
interested in preserving the 3+1d Minkowski invariance, we take the double-Wick rotated solution, along $t$ and $x_6$, giving the background
\bea
ds^2&=&H_4^{-1/2}(r)(-dt^2+d\vec{x}_3^2+ f(r)dx_6^2)+H_4^{1/2}\left(\frac{dr^2}{f}+r^2d\Omega^2_{4}\right)\cr
e^{2\phi}&=&g_s^2H_4^{-\frac{1}{2}}\cr
C_4&=&\frac{1}{g_s}H_4^{-1}dt\wedge dx^1...\wedge dx^4\cr
f&=&1-\left(\frac{r_H}{r}\right)^{3}\cr
H_4&=&1+\alpha_4\left(\frac{r_4}{r}\right)^{3}\label{background}
\eea
This way of breaking susy was  considered in \cite{Aharony:2010mi}, and it gives rise to a semi-infinite
cigar geometry in $r$ over $S^4$, cut off at $r_{min}=r_H$, as well as a cigar geometry in $r$ over $x_6$
(see figure (\ref{cascading})).
We can then consider again (as in \cite{Kachru:2003aw}) cutting off the cigar at a certain $r_{max}$, and gluing another space.
The set-up thus obtained would correspond to KKLT (compactified KS with a susy breaking anti-D3 at the tip). But we want to consider
inflation in the KKLMMT set-up, therefore we will add a sliding D4-brane along the cigar, whose position will be our inflaton.

Considering that the cigar including the T-duality coordinate $x_6$, $(x_6,r,S^4)$ is the space T-dual to the $(r,T^{1,1})$ cigar in KKLT,
which is glued onto a $CY_3$ space $M$. Then $(x_6,r,S^4)$ needs to be glued onto the $CY_3$ space $W$ T-dual to $M$ (note that there is no
simple $U(1)$ in general, so T-duality is generalized).
Strictly speaking, the T-duality picture holds  for the supersymmetric case $f=1$ (or $r_H=0$), but  it will be approximately valid
in the near-extremal case $r_H\ll r_4$. Note that  if we perturb a bit an exact T-duality symmetry by the introduction of a small non-extremality
parameter, we still get an approximate symmetry.
Also, the compactification by gluing of $W$ is supersymmetry preserving, i.e. the gluing does not
generate extra energy. In the near-extremal case, the gluing will not be perfect, so a small additional energy needs to be added to realize it,
but we will argue later that we can neglect it.

We will consider the near-extremal case in the following, for the previous reasons, as well as a number of others that will become
apparent as we analyze the model.

Also, for the T-duality to KS to work, we must add the spiralling brane.
But we will see that we obtain a good inflationary model by just considering the background (\ref{background}), and the effect of the spiralling
brane can be constrained to be small. Therefore we will first analyze the case without the spiral, and then move to the spiral perturbation.
The analysis of the effect of the spiral is nevertheless important, since the spiral guarantees the T-duality with KS, which is known to be
well-defined due to its supersymmetry.

\subsection{Near-extremal D4-brane}

\begin{figure}[htp]
\begin{center}
\includegraphics[width= 90mm]{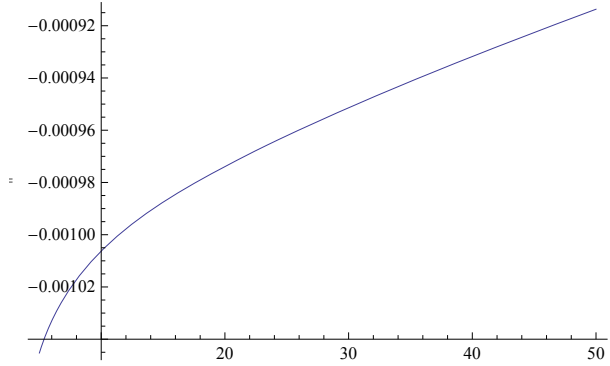}\end{center}
\caption{The potential $V(r)$ in the plateau region, as a function of $r$ for $\beta$=  500. The value of $V(r)$ on the plateau is $1/2\b=0.001$, and
at infinity, $V(r)$ goes to zero.\label{bulkpotential}}
\end{figure}

The  action for a Dp-brane  which includes the DBI term and the CS term takes the form
\be
S_p=-T_p\int e^{-\phi}\sqrt{\det G_{ab}}+\mu_p\int C_{p+1}
\ee
and for a Dp-brane moving in the background of the double-Wick rotated non-extremal Dp-branes one obtains
\be
S=-\frac{T_p}{g_s}\int H_p^{-1}(r)\left[\sqrt{f_p+\frac{H_p(r)}{f_p}g^{\mu\nu}\d_\mu r\d_\nu r}-1\right]\label{Dpaction}
\ee
where $H_p$ and $f_p$ are the analogs of $H_4$ and $f$ given in (\ref {background}).
The corresponding potential for  D4-branes ($p=4$) acting on  a D4-brane sliding in the background (\ref{background}) is
\be
V_4(r)=+\frac{T_4R}{g_s}H^{-1}(r)[\sqrt{f(r)}-1]=
+\frac{T_4R}{g_s}\frac{1}{1+\a_4(\frac{r_4}{r})^3}\left[\sqrt{1-\frac{r_H^3}{r^3}}-1\right]<0\label{barepotential}
\ee
Here $R$ is the radius of the compact $x_6$ without the metric factors,
$r$ is the radial position of the sliding brane and
\bea
(r_4)^3&=&\pi g_sN\a'^{3/2}\cr
\a_4&=&\sqrt{1+(\frac{r_H^3}{2r_4^3})^2}-\frac{r_H^3}{2r_4^3}\label{ralpha}
\eea
so that in the near-extremal case $r_H\ll r_4$, $\a_4\simeq 1$.
The potential is drawn in figure \ref{bulkpotential}.

To analyze this potential, we compute its derivative,
\be
\frac{g_sV'(r)}{T_4R}= 
\frac{3}{2r^4(1+\a_4(\frac{r_4}{r})^3)\sqrt{1-\frac{r_H^3}{r^3}}}\left[\frac{2\a_4r_4^3}{1+\a_4
\frac{r_4^3}{r^3}}\left(1-\frac{r_H^3}{r^3}-\sqrt{1-\frac{r_H^3}{r^3}}\right)+r_H^3\right]
\ee
Denoting $r_H^3/r^3 \equiv x,\a_4r_4^3/r_H^3\equiv \b$, we can check that $V'(r)>0$,
since from $V'(r)=0$ we get the equation $1+\b x +2\b(1-x-\sqrt{1-x})=0$, implying $ \b^2 x^2 -2\b x+(1+2\b)=0$, which has a negative
discriminant, therefore no solution. That means that the potential
increases monotonically from $r=r_H$, where it takes the value
\be
V(r_H)=-\frac{T_4R}{g_s}\frac{1}{1+\a_4(\frac{r_4}{r_H})^3}\simeq -\frac{T_4R}{g_s}\left(\frac{r_H}{r_4}\right)^3,
\ee
to infinity, where it gives zero.
Near $r=r_H$, the potential becomes very steep, with $V'(r=r_H)=\infty$, but the value of the potential stays finite.

Far from the horizon $r_H$, at $r/r_H\gg 1$, we get
\be
V(r)\simeq -\frac{T_4R}{2g_s}\left(\frac{r_H}{r}\right)^3\frac{1}{1+\a_4(\frac{r_4}{r})^3}
\ee
We observe then that even though $V'(r)$ stays always positive, it does in fact stay very close to zero over a large region, if $r_4\gg r_H$,
i.e. if $\b\gg 1$, since then the potential is approximately
\be
V(r)\simeq -\frac{T_4R}{2g_s\a_4}\left(\frac{r_H}{r_4}\right)^3\Big[1+\frac{1}{4}\frac{r_H^3}{r^3}-\frac{r^3}{\a_4r_4^3}\Big]\label{vapprox}
\ee
so is approximately constant over the large region $r_H\ll r\ll r_4$, with $V(r)\simeq V(r_H)/2$.

Note that the potential is negative, but it should really be positive, due to the nonsupersymmetric deformation. The answer to this
puzzle is that the potential we derived is not yet complete. There is also a vacuum energy component, due to the nonsupersymmetric nature of
the background. We are interested in the energy from the point of view of the effective 4d theory in flat space, which means that this energy
needs to be positive since the supersymmetric theory would have zero energy.\footnote{Of course, the total (Casimir) energy of a gravitational space can be
negative even in a nonsupersymmetric set-up. However, we are interested in the energy from the point of view of the effective
theory in flat 4 dimensions, and we know that in that case, supersymmetry requires zero energy. This is exactly the same situation as was
encountered in \cite{Maldacena:2001pb}, so a similar reasoning applies.} This value will depend on the volume of compactification,
but if this volume is large enough the dependence will become negligible, and we will have the vacuum energy of the uncompactified theory, so
we will focus on this in the following.

To compute this energy, one would  need to regularize an integral, corresponding to the gravitational action, over an infinite volume.
Since this regularized gravitational action could be calculated in Euclidean space, and then Wick-rotated, the fact that our solution is
doubly-Wick rotated with respect to the near-extremal solution (we exchanged $t$ with $x_6$) should not matter, and we should get the same result.
Luckily, a simpler procedure for calculating this energy in
the near-extremal case was devised in \cite{Maldacena:2001pb} in order to compute the vacuum energy from the point of
view of the 4d theory in flat space, that guarantees a finite result. The one difference is that in that case, because one was using AdS/CFT,
the theory was defined in the IR of the dual metric (because of the UV/IR relation between gravity and field theory),
i.e. both $g_s$ and the size of the space wrapped by
the branes was defined at small $r$, whereas in our case $g_s$ and $R$ are defined at large $r$, since we have just an effective field theory picture.

A non-extremal solution is obtained by adding to the extremal solution
(representing a set of D-branes), an extra mass $\delta M$ without charge. That $\delta M$ is equivalent to adding $\delta N=\delta M/2$ D-branes
and $\delta N$ anti-D-branes, since a D-brane and an anti-D-brane have mass, but no charge. Therefore we have
\be
\frac{r_H^3}{r_4^3}=\frac{\delta M}{M}=\frac{2\delta N}{N}
\ee
The vacuum energy is the tensional energy of the $2\delta N$ branes wrapping the $x_6$ circle of radius $R$ at coupling $g_s$, in
the same way as in \cite{Maldacena:2001pb}. The result is then
\be
E_0=\frac{T_4R}{g_s}2\delta N=\frac{T_4RN}{g_s}\frac{r_H^3}{r_4^3}\label{constanten}
\ee

The total potential is then the sum of (\ref{barepotential}) and (\ref{constanten}). Note that $E_0=NV(r_H)$, so the resulting potential is very flat
simply due to the large factor of $N$ in the constant part. Also note that we will have an a posteriori check on this calculation in the second
model we will analyze, in eq. (\ref{potsecond}).

Since the background is slightly nonsupersymmetric, in order to compactify by gluing to a CY space we would need to slightly modify the
gluing region, which will create an additional energy, localized near the gluing region. But such an energy will be proportional to $\delta N/N$,
as it should vanish in the supersymmetric case, yet it cannot be proportional to $N$ also, since the gluing region is not drastically modified
by an increase in $N$ (at fixed $\delta N/N$, the space in the absence of the cut-off $r_{max}$ would just scale up, but at fixed cut-off $r_{max}$,
that would be equivalent to scaling the cut-off instead). But at large $r_{max}$, the space is approximately supersymmetric, and if the energy was
 proportional to $N$, by the previous argument, it would mean that the energy could be made infinite by just scaling the cut-off, which is clearly
absurd. Hence the energy is not proportional to $N$.
This means that such a contribution will be much smaller than (\ref{constanten}). It could be comparable to (\ref{vapprox}),
but that does not matter, since it is a constant contribution, so all that matters is the relation to (\ref{constanten}). We can thus
safely ignore it.

\subsection{Adding the spiralling brane}

We now calculate the contribution of the spiralling brane to the potential.

 Since the type IIA configuration is quite complicated-looking, it
is easier to do the calculation in M-theory. We can use the same idea used in  \cite{Kachru:2003sx} to calculate the potential between
the D3 and the anti-D3 in the KS geometry. It was noted that the harmonic function $H(r)$ is harmonic in the transverse space of the background
metric, $dr^2+r^2\tilde g_{ab}dy^ady^b$, and as a result one can calculate the perturbation of the harmonic function due to the supersymmetric
D-brane at a position $r_1$ in the transverse space. We can use then the probe approximation for the anti-D-brane, and compute the interaction
potential via the modification of its DBI action induced by the perturbation of $H(r)$.  We can reverse the logic and
consider the perturbation to
the background given by the anti-D-brane, and calculate its effect on the D-brane. In \cite{Bena:2010ze} it was checked that this gives the
same result in a rather large class of situations. The anti-D-brane would turn the background into a near-extremal one.

In our IIA case, the latter situation  corresponds to having the near-extremal background together with the spiral, and calculate its
effect on the sliding D4-brane. But we can now switch  the points of view yet again, and consider instead the effect of the sliding brane, which would
be supersymmetric for $f=1$, on the background, and calculate the modification of the action of the spiralling brane. This last
switching of points of view is on an even surer footing, since the sliding and spiralling branes are of the same type (branes, not antibranes).

Of course, for this approximation to hold, we should be able to consider the non-extremality as being the effect of only a few anti-D-branes,
i.e. to be in the near-extremal case. We should also note that we will treat the modification of the near-extremal background due to the probe brane
as a modification of the harmonic function $H$ (and possibly $r_H$) only.
This should be correct in the near extremal case only. If the probe brane would be situated at
$r=0$ like the rest, only $H$ would be modified, independent of the non-extremality. For our probe brane at $r>r_H$, there can be other changes besides
the change in $H$ (and possibly $r_H$), but they can only be of higher order in $1/N$,
since the modifications of the other parameters have to be proportional to the
near-extremality parameter $\delta N/N$, whereas the supersymmetric effect (modification of $H$ only) is already proportional to $1/N$, so in total
we must have at least $\delta N/N^2$. We will therefore neglect such modifications in the following. Note that this argument relies on the fact that
the probe is a brane, which would be in a supersymmetric configuration in the absence of the small nonextremality of the background, and thus it
will only modify the parameters of the background ($H$ and $r_H$), but not its form.

Since, as we said, it will be easier to work in M theory, where the spiral is a simple M5-brane,
we consider the M theory uplift of the background,
\bea
ds^2&=&H^{-1/3}[-dt^2+dx_i^2+f(r)dx_6^2+dx_{11}^2]+H^{2/3}[(f(r)^{-1}-1)dr^2\cr
&&+|dv|^2+|dw|^2+dx_7^2]\cr
C_6&=&H^{-1}d^4x\wedge dx_6\wedge dx_{11}\cr
r^2&=&|v|^2+|w|^2+x_7^2
\eea

We have a M5-brane situated at $x_7=0$, so that $r=u$, with
\bea
v&=&u(x_6)e^{i\phi(x_{11})}\cos\a(x_6)\cr
w&=&u(x_6)e^{-i\phi(x_{11})}\sin\a(x_6)
\eea
and winding around $x_6$ many times, spiralling down in $u$ to an $u_{min}$, and then back up.

The fact that in type IIA this describes $p$ D4-branes is realized via the fact that the M5-brane wraps $p$ times around $x_{11}$, so
$\phi(x_{11})=x_{11}/\lambda_p$ (the equation of motion of $\phi$ is $\ddot \phi\equiv \pa^2_{x_{11}}\phi=0$), where
\be
\lambda_p=pg_sl_s=pR_{11}
\ee

The M5-brane action is \cite{Pasti:1997gx,Bandos:1997ui}
\bea
S&=&-T_5\int d^6 x\left[\sqrt{-\det(g_{mn}+\tilde H_{mn})}-\sqrt{-g}\frac{1}{4\d_r a\d^r a}\d_l a H^{*lmn}H_{mnp}\d^pa\right]\cr
&&+\int [C^{(6)}+\frac{1}{2}F\wedge C^{(3)}]\cr
H&=&F-C^{(3)}\cr
\tilde H_{mn}&=&\frac{1}{\sqrt{-(\d a)^2}}H^*_{mnl}\d^l a\cr
H^{*mnl}&=&\frac{1}{3!\sqrt{-g}}\epsilon^{mnlpqr}H_{pqr}\cr
*dC^{(3)}&=&dC^{(6)}+\frac{1}{2}C^{(3)}\wedge dC^{(3)}
\eea
where $a$ is an auxiliary scalar field, needed to avoid explicit breaking of Lorentz invariance, and whose VEV gives for instance
$\d_l a=\delta_{l5}$, but on our background it reduces to only
\be
S=-T_5\int d^6x \left[\sqrt{-\det g_{\mu\nu}^{ind}}-C^{(6)}\right]
\ee
Since we  want to vary the harmonic function, we calculate the action of the M5-brane in the above background as a function of $H$, without
substituting its value, obtaining
\be
-\frac{g_s{\cal L}}{T_4}=H^{-1}\sqrt{1+H(u\dot\phi)^2}\sqrt{f+\frac{H}{f}[(u\a')^2+u'^2]}-H^{-1}
\ee
where $T_4=T_52\pi R_{11}$.
Here prime refers to $\d/\d x_6$ and dot to $\d/\d x_{11}$. The two integrals of motion corresponding to translational invariance in $x_6$ and
$\a$ are then
\bea
&&E=p_u u'+p_\a \a'-{\cal L}=H^{-1}-\frac{H^{-1}\sqrt{1+H(u\dot\phi)^2}f}{\sqrt{f+\frac{H}{f}[(u\a')^2+u'^2]}}\cr
&&J=p_\a=\frac{\sqrt{1+H(u\dot\phi)^2}}{\sqrt{f+\frac{H}{f}[(u\a')^2+u'^2]}}\frac{u^2\a'}{f}\label{EJ}
\eea
We can then eliminate $u'$ and $\a'$ in favor of $E$ and $J$, obtaining
\bea
u\a'&=&\frac{f^2}{1-HE}\frac{J}{u}\frac{1}{\lambda_p}\cr
u'&=&\frac{f}{1-HE}\sqrt{f\left(H^{-1}+u^2-\frac{f J^2}{u^2}\right)-H^{-1}(1-HE)^2}\label{ua}
\eea
where we fixed $\dot\phi=1$ by a rescaling,
\be
u=\lambda_p\bar u\;\;\;\;
u_4=\lambda_p\bar u_4\;\;\;\;
x_6=\lambda_p\bar x_6
\ee
and dropped the bars here and in the following.
Substituting in the action, we obtain
\be
-\frac{g_s{\cal L}}{T_4}=\frac{H^{-2}(1+Hu^2)f}{H^{-1}-E}-H^{-1}
\ee
The potential as a function of the position $u_0$ of the sliding brane is
\be
V(u_0)=-2\lambda_p\int_0^\infty dx_6{\cal L}(u_0,u(x_6))
\ee
since the brane wraps around $x_6$ starting at infinity in $u$, down to $u_{min}$ (corresponding to $x_6=0$) and then back up to infinity.
Substituting the Lagrangian, we obtain
\be
V(r_0)=\frac{2T_4\lambda_p}{g_s}\int_0^\infty dx_6\Big[\frac{H^{-2}f(1+Hu^2)}{H^{-1}-E}-H^{-1}\Big](x_6)\label{potintermed}
\ee

If we put $f=1$ and $E=0$ we return to the supersymmetric case, and then we have \cite{Aharony:2010mi}
\bea
&&u'=\frac{1}{\lambda_p}\sqrt{u^2-\frac{J^2\lambda_p^2}{u^2}};\;\; J\lambda_p\equiv 2\xi^2\cr
&& u^2=2\xi^2\cosh\frac{2x_6}{\lambda_p}=J\lambda_p\cosh\frac{2x_6}{\lambda_p}
\eea
(here and in (\ref{intermedi}) $u$ and $x_6$ are unbarred quantities),
in which case by substituting we obtain an infinite constant,
\be
V=+2T_4\int_0^\infty dx_6 u^2=+2T_4\xi^2\lambda_p \sinh\left(\frac{2x_6}{\lambda_p}\right)|_0^\infty\label{intermedi}
\ee
Since we have no interaction in the supersymmetric case, this is just the rest mass of the spiralling brane, and is divergent since the spiralling
brane is infinite in extent in $x_6$. In a physical case, we must make it finite by regularizing:  we integrate $\int_0^{2\pi Rk}dx_6$,
or correspondingly $\int_{u_{min}}^\Lambda du$, obtaining
\be
V=+2T_4\xi^2\lambda_p\sinh \frac{4\pi Rk}{\lambda_p}\simeq+\lambda_p T_4\Lambda^2
\ee
where the last equality is only valid at large $\Lambda$.

In order to calculate the interaction potential between the spiral and the sliding brane, we calculate the change in $V$ due to the variation in
the harmonic function.  There is also a  variation of $r_H$ of order $1/N$, but we will neglect it at this time, and come back
to it at the end of this subsection.

The harmonic function corresponds to $N$ branes at $r=0$, and the sliding brane adds another one at $r_0$, therefore
\be
H\rightarrow H+\delta H;\;\;\;\;
\frac{\delta H}{H}\simeq \frac{r^3}{N(r-r_0)^3}
\ee
The interaction potential is then
\be
\delta V(r_0)=+\frac{2T_4\lambda_p}{g_s}\int_0^{\infty}dx_6\frac{r^3}{N(r-r_0)^3}H^{-1}\Big\{1-\frac{H^{-1}f}{(H^{-1}-E)^2}\Big[H^{-1}-E(2+Hu^2)\Big]\Big\}
\label{potentia}
\ee
We can check that in the supersymmetric case $f=1$, $E=0$, the interaction potential vanishes.

Substituting $u'$ from (\ref{ua}) in (\ref{potentia}), we finally get for the interaction potential between the sliding brane and the
spiral
\bea
\delta V_{E,J}(r_0, \Lambda)&=&+\frac{2T_4\lambda_p}{g_sN}
\int_{u_{min}(E,J)}^{\Lambda}\frac{du (1-HE)}{f\sqrt{f(H^{-1}+u^2-\frac{f J^2}{u^2})-H^{-1}(1-HE)^2}}\times \cr
&&\times \frac{r^3}{(r-r_0)^3}H^{-1}\Big\{1-\frac{H^{-1}f}{(H^{-1}-E)^2}\Big[H^{-1}-E(2+Hu^2)\Big]\Big\}\cr
&&\label{spiralslide}
\eea
Here $u_{min}$ is the turning point for the spiral, which depends
on $E,J$ by solving the equation $u'(u)=0$, i.e.
\bea
&&\Big[f(H^{-1}+u^2-\frac{f J^2}{u^2})-H^{-1}(1-HE)^2\Big]_{u=u_{min}}=0\Rightarrow\cr
&&\Big(1-\frac{u_H^3}{u_{min}^3}\Big)\Big[\frac{1}{1+\a_4u_4^3/u_{min}^3}+u^2_{min}-\Big(1-\frac{u^3_H}{u_{min}^3}\Big)\frac{J^2}{u_{min}^2}\Big]\cr
&=&\frac{1}{1+\a_4u_4^3/u_{min}^3}\Big(1-\frac{E}{1+\a_4u_4^3/u_{min}^3}\Big)^2\cr
&&
\eea
The potential thus depends on $r_0$ and $\Lambda$ as variables, and $E$ and $J$ as constants parametrizing the solution. Since $\Lambda$ is the
value of $u$ where we glue to the $CY_3$, $\Lambda$ is related to the volume variable.

In the case that $\Lambda$ is sufficiently large so that $H(\Lambda) \simeq 1, f(\Lambda)\simeq 1$, the interaction potential contains a
divergence,
\be
\delta V_{E,J}(r_0,\Lambda)_{div}\sim +\frac{2T_4\lambda_pE}{g_sN(E-1)^2}\left[\frac{\Lambda^2}{2}+3r_0\Lambda+...\right]
\ee

We are interested in the near-extremal case, which means that we can calculate as a perturbation around the susy case $E=0, f=1$.
For small $E$ and $f-1$, we get for the potential (\ref{spiralslide})
\be
\delta V_{E,J}(r_0, \Lambda)\simeq +\frac{2T_4\lambda_p}{g_sN}\int_{\sqrt{J}}^\Lambda\frac{u^3 du}{(u-u_0)^3}
\frac{H^{-1}}{\sqrt{u^2-J^2/u^2}}\Big[-(f-1)+EH^2u^2\Big]
\ee
where we have used the fact that in the susy case (and thus also for the near-extremal case in this order of approximation)
$u_{min}=\sqrt{J}$.

Again if we have a sufficiently large $\Lambda$ such that $H(\Lambda)\simeq 1,f(\Lambda)\simeq 1$, there is the same divergence
\be
\delta V_{E,J}(r_0, \Lambda)_{div}\sim +\frac{2T_4\lambda_p E}{g_sN}\Big[\frac{\Lambda^2}{2}+3u_0\Lambda+...\Big]
\ee

The region of interest for $r_0$
is $r_H\ll r_0\ll r_4$, and there $H(r_0)\simeq \a_4 r_4^3/r_0^3$ and $f(r_0)\simeq 1$, but the integral in $r$ is over a
larger region, where the same does not apply. Therefore we would need to evaluate the integral numerically to get a result.

However, if the region of interest is such that $H(\Lambda)\gg 1$, we can approximate $H(u_0)\simeq \a_4 r_4^3/u_0^3$, giving
\be
\delta V_{E,J}(r_0, \Lambda)\simeq +\frac{2T_4\lambda_p}{g_sN\a_4 r_4^3}\int_{\sqrt{J}}^\Lambda\frac{u^3 du}{(u-u_0)^3}
\frac{1}{\sqrt{u^2-J^2/u^2}}\Big[\frac{E\a_4^2r_4^6}{u}+r_H^3\Big]\label{spiralpot}
\ee
(note that, since we took $H(\Lambda)\gg 1$ to obtain the above, $\a_4^2 r_4^6\gg u^6$, so the first term is $\gg E u^5$, i.e. dominates over
the constant $r_H^3$ term at large $u$, so the leading behavior of the potential, giving the $\Lambda$ behavior, is obtained from
integrating the first term)
and now we see that the contribution of large $u$ goes like $1/\Lambda$, i.e. not only it does not diverges for $\Lambda \rightarrow\infty$, but it actually vanishes.

More importantly, we note that the result is of order $\sim g_sp/N$ with respect to (\ref{vapprox}), so we can consider it a small
correction, and neglect it in the analysis of cosmology.

We now return to the issue of the possible variation of $r_H$ of order $1/N$. Let's write generically $\delta r_H/r_H=\beta/N$.
We can again vary the potential in (\ref{potintermed}) and obtain
\be
\delta V\simeq -\frac{3\beta}{N}\frac{2T_4\lambda_p}{g_s}\int_0^\infty dx_6 \frac{r_H^3}{r^3}\frac{u^2+H^{-1}}{1-HE}
\ee
Again substituting $u'$ from (\ref{ua}) and taking a small $E$ and $f-1$, we obtain
\be
\delta V\simeq -\frac{3\beta}{N}\frac{2T_4\lambda_p}{g_s}u_H^3\int_{\sqrt{J}}^\Lambda\frac{du}{u^3}\frac{u^2+H^{-1}}{\sqrt{u^2-J^2/u^2}}
\ee
We can now check that the integral has a large $\Lambda$ dependence of $1/\Lambda$, and that it is finite at $\sqrt{J}$. Therefore
the perturbation of the potential due to the variation in $r_H$ is of the order of $(g_sp)^3(g_sN)^{2/3}/N$ with respect to
(\ref{vapprox}), therefore again sub-leading, and will be neglected in the analysis of the application to cosmology.

Finally, it is also clear that we can neglect also the modifications due to changes in other quantities besides the harmonic function $H$
and $r_H$, since as
we said those were expected to be even smaller than (\ref{spiralpot}).

\section{A model based on a sliding anti-brane in a supersymmetric (cylinder) background }

In the previous section we have considered the case of a probe brane falling in the background of a doubly Wick rotated
nonextremal D4-brane solution, which itself
could be thought of as being made up of $N+\delta N$ D4-branes and $\delta N$ anti-D4-branes (with a small number $\delta N$ since we
consider the near-extremal case $r_H\ll r_4$), with a spiral brane probe added on the
background.

\begin{figure}[h]
\begin{center}
\includegraphics[width= 80mm]{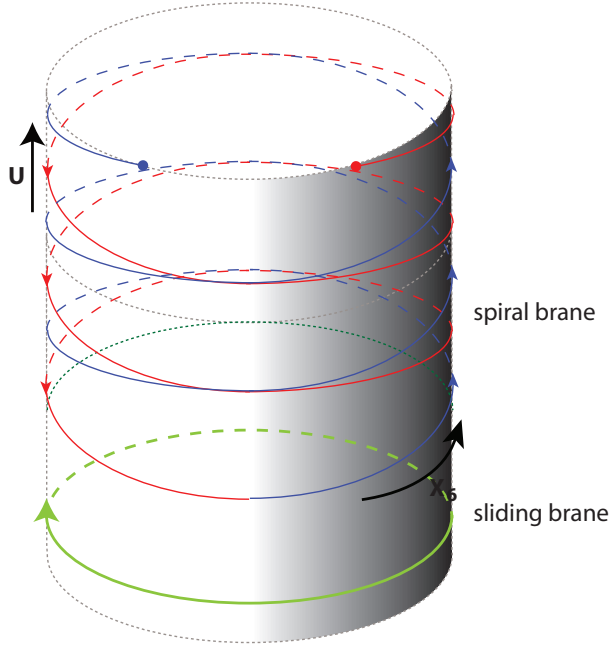}\end{center}
\caption{For compact $x_6$,  the curved $NS5$-brane spirals down the $(x_6, u)$ cylinder
and then climbs back up. The downward (upward) part of the spiral is colored red (blue).\label{cigarspiral}}
\end{figure}

We consider now
the case of a supersymmetric background of $N$ D4-branes and a spiral brane added with a moving anti-brane probe giving the inflaton potential
(see figure (\ref{cigarspiral})).
 If we were able to exactly
describe the case of $\delta N=1$, we could argue as in  \cite{Bena:2010ze}, where it was argued   that exchanging the brane with the antibrane as
probe and part of the background respectively, would in most cases give the same result. In our case, it is less clear, so we will treat here
separately this case. But the upshot is that at $\delta N=1$ we could switch the role of the probe brane and $\delta N=1$ antibrane in the
background and get this second model. A better way of saying this would be that at $\delta N=1$ we cannot think of the antibrane as part
of the background anymore, but instead we should think of it as a probe in a supersymmetric background.

The compactification of the solution happens in exactly the same way as in the previous sections,
since the supersymmetric case is T-dual to the KS background,
as explained in \cite{Aharony:2010mi}. The only difference is that since now we have a supersymmetric case, there is no
 problem anymore in matching the
cylinder with the spiralling brane to the half of CY space. This  will not introduce extra energy at the joining point. On the low $r$ side, the only difference between this case and the previous one is that the space
 terminates at $r=0$ instead of $r=r_H$.

 The Einstein frame metric is

\bea
ds_E^2&=&H^{1/2}H^{-1/8}[dr^2+r^2d\Omega_4^2+Hdx_6^2]+...\cr
&\sim&r^{-9/8}[dr^2+r^2d\Omega_4^2+r^3dx_6^2]+...\cr
&\sim& dy^2+y^2d\Omega_4^2+y^{30/7}dx_6^2+...
\eea
so  the space terminates at $y=16/7r^{7/16}=0$. Of course, the solution above is singular at $r=0$, so there will be corrections to the
geometry, but
we will still have a space that terminates, so the extra dimensions are truly compactified by this construction.

The calculation of the potential
is the same as before, just putting $r_H=0$ and changing the sign of the CS term because we work with an anti-brane probe, i.e.
chaging the $-1$ to $+1$ in (\ref{barepotential}), obtaining
\be
V_4(r)=\frac{2T_4R}{g_s}\frac{1}{1+\frac{r_4^3}{r^3}}\label{potsecond}
\ee
Now, since in the absence of the antibrane probe we have a supersymmetric model, we don't have any constant term to add. However, it is
satisfying to observe that the constant term in the above potential, $V_4(\infty)$, equals the constant term we assumed in the previous section
for $\delta N=1$, as it should, since we argued that the two models are related just by the exchange of a $\delta N=1$ brane part of the
background and a brane probe.

The potential now varies between $r=0$ where it is $V(0)=0$, with the derivative giving
\be
V_4'(r)=\frac{6T_4R}{g_sr}\frac{r_4^3/r^3}{1+r_4^3/r^3}
\ee
which goes to infinity at $r=0$, and at infinity the potential flattens out to $V(\infty)=2T_4R/g_s$.

We next turn to the computation of the correction to  the potential due to the spiralling brane.
We would like to use the same logic as was used above for the sliding brane on the non-supersymmetric cigar background, namely, instead of computing the impact of the spiral on the sliding anti-brane we will aim to compute the reverse, namely the  impact of the anti-brane on the spiral, and
moreover we do the computation using the effect of the probe on the background.

Had it been a sliding brane rather than an anti-brane  the contribution to the potential of the spiralling brane would have been vanishing as we have seen in (\ref{potentia}) when we take $f=1$ and $E=0$. This would follow from a cancelation between the DBI and CS contributions to the potential. Now for the potential acting on a sliding  anti-brane the contribution of the DBI term is the same on the sliding brane but that of the CS term has an opposite sign. Thus altogether the contribution of the spiralling brane to the potential acting on the sliding anti-brane will be twice the contribution of the DBI  to the potential acting on a sliding brane, namely
\bea
&&\delta V(r_0)=\frac{4T_4}{g_s}\int_0^{\infty}dx_6\frac{r^3}{N(r-r_0)^3}H^{-1}\cr
 &&=\frac{2T_4\lambda_p}{g_sN}\int_{\sqrt{J}}^\Lambda\frac{u^3 du}{(u-u_0)^3}
\frac{H^{-1}}{\sqrt{u^2-J^2/u^2}}\cr
&&\simeq \frac{4T_4\lambda_p}{g_sN\a_4 r_4^3}\int_{\sqrt{J}}^\Lambda\frac{u^6 du}{(u-u_0)^3}
\frac{1}{\sqrt{u^2-J^2/u^2}}
\label{potentialu}
\eea
where in the last equality we took the near horizon harmonic function $H \simeq \alpha_4\left(\frac{r_4}{u}\right)^{3}$. We see that in this
case, we get a very strong $\Lambda$ dependence (volume dependence) at large $\Lambda$, namely $\propto \Lambda^3$.
If we take $H(u)\sim 1$ instead, we get a result which behaves like $\ln \Lambda $, a much milder dependence. In any case, we cannot
have a too large volume in this case. But again the parametric dependence of the potential correction
(\ref{potentialu}) (at least in the $H(u)\sim 1$ case) is $\sim g_sp/N$ with respect to the leading term (\ref{potsecond}).

Hence the impact of spiral may be neglected as before. We should also mention however that the calculation of the impact of the spiral has a
potential caveat: we calculate the interaction of the two probes (spiral and sliding) via the modification of the background, but we assume that
the only modification when changing a brane to an antibrane is the change in relative sign of the CS and DBI terms, but there is no backreaction
on the first probe brane.
If we  do the interaction in flat space, this approximation would not be valid, however because the calculation is done via the effect on a nontrivial
background, it is likely  to be valid. In any case, we only wanted to show the correction is small, we will not use the form of (\ref{potentialu})
in the following.


This calculation then looks promising for cosmology, due to the flatness at large $r$,
but we will see in the next section that we have the usual problem that for the cosmology agrees with experiments only for non-generic initial
conditions.

\section{Cosmology of the model and experimental constraints}

Given a potential with a sufficiently flat region, we can use the standard formulas for inflation to determine the constraints on the
parameters. An issue we  want to examine is the  generality of initial conditions. We therefore express the formulas in terms of the
fundamental Planck scale $m$, $\a'$ and the 4d Planck scale $M_P$. The relations of these scales are
\bea
&&2\kappa_{10}^2=(2\pi)^7\a'^4=m^{-8}\cr
&&M_P=m^4\sqrt{RV_5}
\eea
where $V_5$ is the volume of the 5d space, including the metric in front of $dx_6^2$, and $\kappa_{10}$ is the 10d Newton constant. The 4-brane
tension is
\be
T_4=\frac{1}{2\sqrt{\pi \a'}\kappa_{10}}=\frac{m^4}{\sqrt{2\pi\a'}}
\ee

\subsection{The  model of a sliding brane on a cigar background}

The potential before adding the spiral is given by (\ref{barepotential}) plus (\ref{constanten}), but it is not yet written in terms of the
canonical scalar $\phi$.

From (\ref{Dpaction}), the kinetic term for $r$ is
\be
-\frac{T_4R}{g_s f^{3/2}(r)}\frac{\d_\mu r\d^\mu r}{2}
\ee
Since we are interested in the regime $r/r_H\ll 1$, we can put $f(r)\simeq 1$ in the denominator and define the canonical scalar as
$\phi=r\sqrt{T_4R/g_s}$, and for $r_H$ and $r_4$ corresponding $\phi_H$ and $\phi_4$. We obtain on the plateau
\be
V(\phi)\simeq \frac{T_4R}{g_s}\left(\frac{\phi_H}{\phi_4}\right)^3\Big[N-\frac{1}{2\a_4}\left(1+\frac{\phi_H^3}{4\phi^3}-\frac{\phi^3}{\a_4\phi_4^3}
\right)\Big]\label{canpotapp}
\ee

In order to get a good model of inflation, we need that the slow-roll parameters $\epsilon$ and $\eta$ defined by
\bea
\epsilon&\equiv &\frac{M_p^2}{2}\left(\frac{V'}{V}\right)^2\cr
\eta&=&M_P^2\frac{V''}{V}
\eea
are much smaller than 1, since they give the spectral index of primordial scalar fluctuations, $P^s_\delta(k)\sim k^{n_s-1}$,
\be
n_s-1\equiv \frac{d\ln P^s_\delta (k)}{d\ln k}=-6\epsilon+2\eta
\ee
which is almost exactly flat ($n_s=1$). We also need that the number of e-foldings during inflation, given in terms of the potential by
\be
{\cal N}\equiv  \int \frac{H_a}{\dot\phi}d\phi=\int_{\phi_{end}}^{\phi_i}\frac{V}{M_P^2V'}d\phi=\frac{1}{M_P}\int_{\phi_{end}}
^{\phi_i}\frac{d\phi}{\sqrt{2\epsilon}}
\ee
(where $H_a$ is the Hubble  constant)
is larger than about 60, the COBE normalization constraint for the value of the potential at the end of inflation,
and finally that we get a sufficiently large reheating temperature $T_H$ at the end of inflation, which generically
requires that the potential energy at the start of inflation be not too far below the Planck scale. However, depending on the model, in particular on
the couplings to matter, we can have smaller reheating temperature, so we will not use the above for any constraint.

From the approximate potential (\ref{canpotapp}) over the flat region $r_H\ll r\ll r_4$, we get
\bea
\epsilon&\equiv& \frac{1}{2}\left(M_P\frac{V'}{V}\right)^2
\simeq \frac{1}{2}\left[\frac{3M_P}{2N\phi}\left(\frac{1}{4}\frac{\phi_H^3}{\phi^3}+\frac{\phi^3}{\phi_4^3}\right)\right]^2\cr
\eta&\equiv& M_P^2\frac{V''}{V}=\frac{3M_P^2}{2N\phi^2}\left(-\frac{\phi_H^3}{\phi^3}+2\frac{\phi^3}{\phi_4^3}\right)
\label{epsilon}
\eea
which can be very small, even if $M_P/\phi$ is generic or even large (small $\phi$). All we needed for this result was $\phi_H\ll \phi\ll \phi_4$,
which is a generic case, since we saw that we needed $\phi_H$ to be small for our construction to be valid (one could argue whether such a $\phi_H$
is natural or not) and $\phi_4/M_P$ is generically large, as seen from (\ref{phi4MP})
(since $N$ is large and $g_s$ is small, we would need large $V_5$ to obtain small $\phi_4/M_P$, which is possible, but non generic).
Therefore a generic $\phi$, of order $M_P$, will fall within the plateau regime.

Experimentally, a red spectrum ($n_s-1<0$) is preferred (see for instance \cite{Baumann:2009ds}). Since from the above we have generically
$\epsilon\ll |\eta|$, the condition for a red spectrum is $\eta<0$, or $\phi<\sqrt{2^{-1/3}\phi_H\phi_4}$.


For the number of e-folds we obtain
\be
{\cal N}\simeq \frac{2N}{3}\int_{\phi_{end}}^{\phi_{in}}\frac{d\phi}{M_P}\frac{\phi/M_P}{\frac{\phi_H^3}{4\phi^3}+\frac{\phi^3}{\phi_4^3}}
\ee
Assuming we can neglect $\phi^3/\phi_4^3$ with respect to the first term over the period of interest, we finally obtain
\be
{\cal N}=\frac{8N}{15}\frac{\phi_{in}^5-\phi_{end}^5}{M_P^2\phi_H^3}\label{efolds}
\ee
which can easily be made large enough.

The COBE normalization constraint states that the magnitude of the scalar fluctuations during inflation (specifically, at horizon exit, but since
we are on the plateau, what is constrained is the plateau value),
\be
<\Delta \phi>=\frac{H_e}{2\pi}
\ee
must give rise to the observed CMBR fluctuations, i.e. must equal $\sqrt{2\epsilon}10^{-5}M_P$ ($\delta\rho/\rho=H_e/(2\pi \sqrt{2\epsilon}M_P)$).
This puts a constraint for the magnitude of the
potential on the  inflation plateau $V_p$, since $2\pi V_p/3={H_e}^2M_P^2$, giving
\be
V_p\sim 12\pi \epsilon \times 10^{-10}M_P^4
\ee
Since we have
\be
V_p=N\left(\frac{T_4R}{g_s}\right)\left(\frac{\phi_H}{\phi_4}\right)^3
\ee
we can think of this as a constraint on $(T_4R/g_s)$ once $N$ and $\phi_H^3/\phi_4^3$ are fixed by (\ref{epsilon}) and (\ref{efolds}).

{\bf Reheating and relaxing to zero potential}

At $r=r_H$ we have $V'(r\sim r_H)\sim 1/\sqrt{f}\rightarrow\infty$, as we saw. It is true that then the kinetic term for $r$ must be put in the
canonical form, however from (\ref{Dpaction}) we see that at $r\simeq r_H$ we don't have a nonlinear sigma model, but rather we have a
nonstandard and divergent kinetic term $\sim \sqrt{(\d \phi)^2/(r-r_H)}$, signifying that the effective description in terms of a single
scalar field is breaking down. Before that however, the derivative of the scalar potential will become large, and inflation will end, so the
breakdown region corresponds to the where reheating should take place.

Therefore in this region string corrections should
become important, at least in the interesting case when the number of anti-branes making the background nonextremal, $\delta N$,
is of order 1. In this case we know that there is not much reason to consider $\delta N$ as part of the background, while one brane is kept as a probe,
at least not when the probe is close to the antibranes (located near $r=0$, or $r=r_H$). Instead a better description would involve antibranes
at $r=0$, which would annihilate with the sliding brane, generating string corrections. So the slope of the potential at $r_H$,
as well as its depth at $r_H$, could be corrected anyway. In any case, the nonsusy background will be unstable, so some time after the probe brane hits
$r_H$ we should have a decay process, made favorable also by this collision, and we should decay to a supersymmetric vacuum of zero potential.

This process will reheat the Universe, through decay into matter modes of the energy released in the fall to the susy vacuum, as usual
in brane-antibrane inflation, see for instance \cite{Burgess:2001fx}.
In principle, two possible scenarios can occur: the potential could have zero
slope at $r=r_H$, and we could have standard reheating through oscillations. Or, in view of the above, more likely is the usual brane-antibrane
case, of a very steep potential, giving rise to preheating (see \cite{Allahverdi:2010xz} for a review).
But as usual, this is a complicated nonperturbative process.
If the potential is steep enough and the coupling to matter large enough,
we expect to generate a large enough reheating temperature, though we will not attempt a further description.

In conclusion, in this model we can easily satisfy experimental constraints with generic initial conditions for $\phi$.

Of course, the usual caveat present in \cite{Kachru:2003sx} still applies in the same form, since the supersymmetric case is T dual to the one
considered there. When we try to stabilize the volume modulus, generically we will spoil the slow-roll conditions of the potential.

\subsection{{The  model of a sliding  anti-brane on a cylinder  background}
}

In this case, $f=1$, so the kinetic term has only a constant rescaling, i.e. $\phi=r\sqrt{T_4R/g_s}$, giving
\be
V_4(\phi)=\frac{2T_4R}{g_s}\frac{1}{1+\frac{\phi_4^3}{\phi^3}}
\ee
The slow roll parameters $\epsilon,\eta$ are
\bea
\epsilon&=&\frac{1}{2}\left[\frac{3M_P}{\phi} \frac{\phi_4^3/\phi^3}{1+\phi_4^3/\phi^3}\right]^2\cr
\eta&=&-\frac{M_P^2}{\phi^2}\frac{\phi_4^3/\phi^3(32+30\phi_4^3/\phi^3)}{(1+\phi_4^3/\phi^3)^2}
\eea
But $\phi_4$ in Planck units is
\be
\frac{\phi_4}{M_P}=\left[\frac{\pi^{1/4}}{2^{3/4}\sqrt{g_s}}\right]^{1/3}\sqrt{\frac{\sqrt{\a'}N^{2/3}}{m^4V_5}}\label{phi4MP}
\ee
We then see that in order to have a small $\epsilon$ and $\eta$ we can either have $\phi\gg M_P$ which is non-generic (and difficult to
obtain, as type IIB examples show) or,
if instead $\phi\sim M_P$, we need $\phi\gg \phi_4$, which means we need $\phi_4/M_P\ll 1$. We see that this latter case is only possible if $V_5m^4/(\sqrt{\a'}N^{2/3})\gg 1$,
which is possible, but is again non-generic. If we allow for either of these two non-generic cases however, we can also get a large enough number of
e-folds, since
\be
{\cal N}=\int_{\phi_{end}}^{\phi_{in}}d\phi\frac{\phi^4(1+\phi_4^3/\phi^3)}{3M_P^2\phi_4^3}\simeq \frac{\phi_{in}^5-\phi_{end}^5}{3M_P^2\phi_4^3}
\ee
where in the second equality we have assumed $M_P\sim \phi\gg \phi_4$. The COBE normalization gives now
\be
V_f\sim \frac{2T_4R}{g_s}\sim 12\pi \epsilon \times 10^{-10}M_P^4
\ee

We also note now that, since $\epsilon>0$ by definition and $\eta<0$ in this case, we always have the experimentally preferred red spectrum.

Reheating is simpler in this model, since now we already know that inflation ends at $\phi=0$, where $V(0)=0$, where all the energy goes into
matter modes, and moreover
\be
\frac{ dV_4}{d\phi}(\phi)=\frac{6T_4R}{g_s\phi}\frac{\phi_4^3/\phi^3}{1+\phi_4^3/\phi^3}
\ee
so the slope of the potential in canonical scalar variable blows up at $\phi=0$, and we have the usual preheating scenario. Note that in this case,
due to the supersymmetry of the background, it is hard to see how there could be string corrections to the potential near $r=0$, so the fact that
the derivative of the potential blows up seems robust.

\section{Conclusions}

In this paper we have analyzed  type $II_A$ inflation scenarios based on the MQCD model \cite{Aharony:2010mi} . The supersymmetric case, T dual to a compactified KS model,
is a Wick-rotated D4-brane with a spiral 5-brane added, and compactified by gluing a CY space. For a model of inflation we considered two
cases.

For a near-extremal background with a moving D4-brane probe, we obtained a flat enough potential, with slow-roll conditions and
normalization which are  easy to satisfy for generic initial conditions. We also saw that the interaction of the probe brane with the spiral is negligible
in the final potential. Reheating in this model is harder than usual to analyze because the small $r$ region is unreliable in this probe brane
approximation, especially if $\delta N=1$ in the background. We also argued that the additional energy due to the gluing of the CY, and
modifications of the near-extremal background due to the probe brane in other quantities than $H$ are small, but it would be useful to find a
way to calculate them directly.

In the second model, we considered a supersymmetric background, with  an anti-D4 probe brane. Then the gluing of the CY doesn't introduce extra
energy, but there are several potential caveats for interaction of the probe with the spiral. We have seen however that likely we can again ignore
this correction to the potential. The model can produce slow-roll inflation, and reheating is
easier to describe, but the initial conditions for inflation are non-generic.

We note that the final inflationary models obtained were not simply related to the KKLMMT one. The supersymmetric version of our models was related
by T-duality, but the supersymmetry breaking was enough to guarantee new results for cosmology. It will be interesting to study further
the physical implications of these models.

In both of these models, we have the usual problem of that the  stabilization of the volume modulus generically
spoils  the slow-roll,  in a similar manner to  the  KKLMMT model. Since \cite{Kachru:2003sx}, a lot of work has been done on the problem of moduli stabilization in the context of brane sliding
on a throat stabilized by fluxes, for example
\cite{Baumann:2006th,
Baumann:2007np,
Baumann:2007ah,
Baumann:2008kq,
Baumann:2009qx,
Baumann:2010sx,
Baumann:2010ys,
Agarwal:2011wm}.
For some recent reviews of string inflation with a more complete list of references,
see \cite{McAllister:2007bg,Baumann:2009ni,Burgess:2011fa}. Generically, the stabilization of the moduli gives a large mass to the
inflaton (brane position), spoiling slow-roll, but there are extra contributions which in principle could cancel this
(see e.g. \cite{Baumann:2008kq}) in a fine-tuned manner. Otherwise, one would have to rely on fine-tuned initial conditions for inflation.
The problem is still that in order for this to work, one would have to consider a comprehensive modular potential, including all
possible contributions, and that is
a very difficult task. Progress in this direction was realized in \cite{Baumann:2009qx,Baumann:2010ys}. In a general potential, there are too
many contributions, but in \cite{Agarwal:2011wm}, a statistical approach revealed that in a small (fine-tuned) set of potential coefficients can
give rise to inflation, independent on initial conditions.
However, here we have not attempted to address the issue of moduli stabilization, since as described above it is a complicated issue and we leave it to future investigation. Having saying that, let us speculate about a possible scenario  for  the  stabilization  of the volume modulus. In analogy with the mechanism in the KKLT  model, one can  contemplate adding  a D8 brnae that wraps the  circle as well as the $S^4$. Such a brane may yield a gaugino condensation and thus produce a volume-stabilizing potential. The addition of such a brane should not spoil the slow-roll properties of the inflaton potential since it will be suppressed  by powers of $1/N$ in a similar manner to the suppression of the contribution to the potential from the spiral brane discussed above.

{\bf Acknowledgements}. We thank Ofer Aharony for many useful comments on the project and for a critical reading of the manuscript, to Oleg Lunin for a discussion  and to
Nissan Itzhaki for useful suggestions on the manuscript. The research of HN is supported in part by CNPQ grant 301219/2010-9. The work of JS was supported in part by
 a centre of excellence supported by the Israel
Science Foundation (grant number 1468/06), by a grant (DIP H52) of the German
Israel Project Cooperation, and  by a BSF grant.

\newpage

\bibliographystyle{utphys}
\bibliography{cosmoMQCD}

\Comment{

}

\end{document}